\documentclass[12pt,notitlepage]{article}
\usepackage{amsmath}

\newtheorem{theorem}{Theorem}

\newtheorem{corollary}[theorem]{Corollary}

\newtheorem{definition}[theorem]{Definition}

\newtheorem{lemma}[theorem]{Lemma}

\newtheorem{proposition}[theorem]{Proposition}

\newenvironment{proof}[1][Proof]{\noindent\textbf{#1.} }{\ \rule{0.5em}{0.5em}}
\input{tcilatex}
\begin{document}

\title{A Normal Form Game Model of Search and Pursuit\\
	\large To appear in the forthcoming edition of \textit{Advances in Dynamic Games and Applications}}
\author{Steve Alpern and Viciano Lee}
\date{}
\maketitle

\begin{abstract}
Shmuel Gal and Jerome Casas have recently introduced a game theoretic model
that combines search and pursuit by a predator for a prey animal. The prey
(hider) can hide in a finite number of locations. The predator (searcher)
can inspect any $k$ of these locations. If the prey is not in any of these,
the prey wins. If the prey is found at an inspected location, a pursuit
begins which is successful for the predator with a known capture probability
which depends on the location. We modify the problem so that each location
takes a certain time to inspect and the predator has total inspection time $%
k.$ We also consider a repeated game model where the capture probabilities
only become known to the players over time, as each successful escape from a
location lowers its perceived value capture probability.

\end{abstract}

\section{Introduction}

Traditionally, search games and pursuit games have been studied by different
people, using different techniques. Pursuit games are usually of perfect
information and are solved in pure strategies using techniques involving
differential equations. Search games, on the other hand, typically require
mixed strategies. Both Pursuit and Search games were initially modelled and solved by Rufus Isaacs in his book (1965). The first attempt to combine these games was the elegant
paper of Gal and Casas (2014). In their model, a hider (a prey animal in
their biological setting) begins the game by choosing among a finite set of
locations in which to hide. The searcher (a predator) then searches (or
inspects) $k$ of these locations, where $k$ is a parameter representing the
time or energy available to the searcher. If the hiding location is not
among those inspected, the hider wins the game. If the searcher does inspect
the location containing the hider, then a pursuit game ensues. Each location
has its own capture probability, known to both players, which represents how
difficult the pursuit game is for the searcher. If the search-predator
successfully pursues and captures the hider-prey, the searcher is said to
win the game. This is a simple but useful model that encompasses both the
search and the pursuit portions of the predator-prey interaction.

This paper has two parts. In the first part, we relax the assumption of Gal
and Casas that all locations are equally easy to search. We give each
location its own search time and we give the searcher a total search time.
Thus he can inspect any set of locations whose individual search times sum
to less than or equal to the searcher's total search time, a measure of his resources or
energy (or perhaps the length of daylight hours, if he is a day predator).
We consider two scenarios. The first scenario concerns $n$ hiding locations, in which the search time at each location is inversely proportionate with the capture probability at that location. In the second, we consider that
there are many hiding locations, but they come in only two types,
identifiable to the players. Locations within a type have the same search
time and the same capture probability. There may be any number of locations
of each type.

The second part of the paper relaxes the assumption that the players know
the capture probability of every location precisely. Rather, we assume that
a distribution of capture probabilities is known. The players can learn
these probabilities more precisely by repeated play of the game. We analyze
a simple model with only two locations and two periods, where one location
may be searched in each period. While simple, this model shows how the
knowledge that the capture probabilities will be updated in the second
period (lowered at a location where there was a successful escape) affects
the optimal play of the game.

\section{Literature Review}

An important contribution of the paper of Gal and Casas discussed in the
Introduction is the analysis involves finding a threshold of locations beyond which the searcher can inspect. If this is sufficiently
high, for example if he can inspect all locations, then the hider adopts the
pure strategy of choosing the location for which the probability of
successful pursuit is the smallest. On the other hand, if $k$ is below this
threshold (say $k=1),$ the hider mixes his location so that the probability
of being at a location multiplied by its capture probability (the
desirability of inspecting such a location) is constant over all locations.

The paper of Gal and Casas (2014) requires that the searcher knows his
resource level (total search time) $k.$ In a related but not identical model
of Lin and Singham (2016) it is shown that sometimes the optimal searcher
strategy does not depend on $k.$ This paper is not directly related to our findings but reader may find it useful to know the distinction between this paper and ours.

Alpern, Gal, and Casas (2015) extended the Gal-Casas model by allowing
repeated play in the case where the searcher chose the right location but the pursuit at this
hiding location is not successful. They found
that the hider should choose his location more randomly when the pursuing
searcher is more persistent.

More recently, Lidbetter and Hellerstein (2017) introduced an algorithm
similar to that of the \textit{fictitious} play where the searcher
recursively updates his optimal strategy after knowing the response of the
opponent's. They apply this technique to games similar to those we consider
here. Their algorithm is likely to prove a powerful technique for solving
otherwise intractable search games.

More generally, search games are discussed in Alpern and Gal (2006) and
search and pursuit problems related to robotics are categorized and
discussed in Chung, Hollinger and Isler (2011).

\section{Single Period Game with General Search Times}

Consider a game where the searcher wishes to find the hider at one of $n$
locations and then attempt to pursue and capture it, within a limited amount
of resources denoted by $k$. Each location $i$ has two associated
parameters: a \textit{search time} $t_{i}$ required to search the location
and a \textit{capture probability} $p_{i}>0$ that if found at location $i$
the searcher's pursuit will be successful. Both $t_{i}$ and $p_{i}$ are
known to the searcher and the hider.

The game $G\left( n,t,p,k\right) ,$ where $t=\left( t_{1},\dots
,t_{n}\right) $ and $p=\left( p_{1},\dots ,p_{n}\right) $ represent the time
and capture vectors, is played as follows. The hider picks a location $i\in
N\equiv \left\{ {1,2,...,n}\right\} $ in which to hide. The searcher can
then inspect search locations in any order, as long as their total search
time does not exceed $k$. The searcher wins (payoff = 1) if he finds and
then captures the hider; otherwise the hider wins (payoff = 0). We can say
that this game is a constant sum game where the value $V=V(k)$ is the
probability that the predator wins with given total search time $k$. \newline
A mixed strategy for the hider is a distribution vector $h\in $ $H,$ where 
\begin{equation*}
H=\left\{ h=\left( h_{1},h_{2},\dots ,h_{n}\right) :h_{i}\geq
0,\sum_{1}^{n}h_{i}=1\right\} .
\end{equation*}

A pure strategy for the searcher is a set of locations $A\subset N$ which
can be searched in total time $k.$ His pure strategy set is denoted by $%
a(k), $ where

\begin{equation*}
a(k)=\{A\subset N:T(A)\equiv \sum_{i\in A}t_{i}\leq k\}.
\end{equation*}%
The statement above simply states that a searcher can inspect any set of
locations for which the total search time does not exceed his maximum search
time $k.$ A mixed search strategy is a probabilistic choice of these sets.%
\newline

\textit{The payoff} $P$ from the perspective of the maximizing searcher is
given by 
\begin{equation*}
P(A,i)=%
\begin{cases}
p_{i}\quad \text{if}\quad i\in A,\text{ and} \\ 
0\quad \text{if}\quad i\notin A.%
\end{cases}%
\end{equation*}

As part of the analysis of the game, we may wish to consider the best
response problem faced by a searcher who knows the distribution $h$ of the
hider. The "benefit" of searching each location $i$ is given by $%
b_{i}=h_{i}~p_{i}$, the probability that he finds and then captures the
hider (prey). Thus when $h$ is known, the problem for the searcher
essentially is to choose the set of locations $A\in \alpha \left( k\right) $
which maximizes $b(A)=\sum_{i\in A}b_{i}$. This is a classic Knapsack
problem from the Operations Research literature (A seminal book of the Knapsack problem is by Kellerer, Pferschy and Pisinger (2004)). The \textit{objects} to be
put into the \textit{knapsack} are the locations $i.$ Each has a `weight' $%
t_{i}$ and a benefit $b_{i}$. He wants to fill the knapsack with as much
total benefit subject to a total weight restriction of $k.$

The knapsack approach illustrates a simple domination argument: the searcher
should never leave enough room (time) in his knapsack to put in another
object. However to better understand this observation, we show the definition of \textit{Weakly dominant} below

\begin{definition}
	Strategy X weakly dominates strategy Y iff (1) X never provides a lower payoff than Y against all combinations of opposing strategies and (2) there exists at least one combination of strategies for which the payoffs for X and Y are equal.
\end{definition}

  Having stated this, we write this simple observation as follows.

\begin{lemma}
\label{domination lemma}Fix $k.$ The set $A\in \alpha \left( k\right) $ is
weakly dominated by the set $A^{\prime }\in \alpha \left( k\right) $ if $%
A\subset A^{\prime }$ and there is a location $j\in A^{\prime }-A$.
\end{lemma}

\begin{proof}
If $i$ is in both $A$ or $i$ is not in $A^{\prime },$ then $P\left( A,i\right)
=P\left( A^{\prime },i\right) .$ If $i\in A^{\prime }-A$ then $P\left(
A^{\prime },i\right) =p_{i}>0=P\left( A,i\right) .$
\end{proof}

\subsection{An example}

To illustrate the general game we consider an example with $n=4$ locations.
The search times are given by $t=\left( 5,3,4,7\right) $ and the respective
capture probabilities are given by $p=\left( .1,.2,.15,.4\right) .$ In this
example it is easiest to name the locations by their search time, so for
example the capture probability at location $7$ is $0.4.$ The searcher has
total search time given by $k=7,$ so he can search any single location or
the pair $\left\{ 3,4\right\} .$ The singleton sets $\left\{ 3\right\} $ and 
$\left\{ 4\right\} $ are both dominated by $\left\{ 3,4\right\} .$ We put
the associated capture time next to the name of each location. Thus the
associated reduced matrix game is simply%
\begin{equation*}
\begin{tabular}{lllll}
A\TEXTsymbol{\backslash}location & 5~$\left( .1\right) $ & 3~$\left(
.2\right) $ & 4~$\left( .15\right) $ & 7 $\left( .4\right) $ \\ \cline{2-5}
$\left\{ 5\right\} $ & \multicolumn{1}{|l}{.1} & \multicolumn{1}{|l}{0} & 
\multicolumn{1}{|l}{0} & \multicolumn{1}{|l|}{0} \\ \cline{2-5}
$\left\{ 7\right\} $ & \multicolumn{1}{|l}{0} & \multicolumn{1}{|l}{0} & 
\multicolumn{1}{|l}{0} & \multicolumn{1}{|l|}{.4} \\ \cline{2-5}
$\left\{ 3,4\right\} $ & \multicolumn{1}{|l}{0} & \multicolumn{1}{|l}{.2} & 
\multicolumn{1}{|l}{.15} & \multicolumn{1}{|l|}{0} \\ \cline{2-5}
\end{tabular}%
\end{equation*}%
Solving the matrix game using online solver (Avis, et.al. (2010)) shows that the prey hides in the four locations with
probabilities $\left( 12/23,0,8/23,3/23\right) $ while the searcher inspects 
$\left\{ 5\right\} $ with probability $12/23,$ $\left\{ 7\right\} $ with
probability $3/23,$ and $\left\{ 3,4\right\} $ with probability $8/23.$ The
value of the game, that is, the probability that the predator-searcher finds
and captures the prey-hider, is $6/115.$ Our approach in this paper is not
to solve games in the numerical fashion, but rather to give general
solutions for certain classes of games, as Gal and Casas did for the games
with $t_{i}=1.$

\subsection{The game with $t_{i}$ constant}

Choosing all the search times $t_{i}$ the same, say $1,$ we may restrict $k$
to integers. This is the original game introduced and solved by Gal and
Casas (2014). Since the $t_{i}$ are the same, we may order the locations by
their capture probabilities, either increasing or decreasing. Here we use
the increasing order of the original paper. Clearly if $k=1$ the hider will
make sure that all the locations are equally good for the searcher $%
(p_{i}h_i) =$constant) and if $k=n$ the hider knows he will be
found so he will choose the location with the smallest capture probability
(here location $1$). The nice result says that there is a threshold value
for $k$ which divides the optimal hiding strategies into two extreme types.

\begin{proposition}[Gal-Casas]
Consider the game $G\left( n,t,p,k\right) $ where $t_{i}=1$ for all $i$ and
the locations are ordered so that $p_{1}\leq p_{2}\leq \dots \leq p_{n}.$
Define $\lambda =\tsum\limits_{i=1,n}1/p_{i}.$ The value of this game is
given by $\min \left( k\lambda ,p_{1}\right) .$ If $k<p_{1}/\lambda $ then
the unique optimal hiding distribution is $h_i =\lambda /p_{i},$
$i=1,\dots ,n.$ If $k\geq p_{1}/\lambda $ then the unique optimal hiding
strategy is to hide at location $1.$
\end{proposition}

\subsection{The game with $t_{i}=i,$ $p_{i}$ decreasing, $k=n$ odd.}

We now consider games with $t_{i}=i$ and $p_{i}$ decreasing. In some sense
locations with higher indices $i$ are better for the hider in that they take
up more search time and have a lower capture probability. Indeed if the
searcher has enough resource $k$ to search all the locations $%
(k=\tsum\nolimits_{i=1,n}t_{i}=n\left( n+1\right) /2)$ then of course the
hider should simply hide at location $n$ and keep the value down to $p_{n}.$
 Note that if $k<n,$
the hider can win simply by hiding at location $n,$ which takes time $%
t_{n}=n $ to search. We give a complete solution for the smallest nontrivial
amount of resources (total search time) of $k=n.$ Let us first define the following two variables which will be widely used in our main result.\\

\begin{center}
	
$S(p)=\sum_{j=m+1}^{2m+1}1/p_{j};$ \space $\bar{h}_j =1/\left( p_{j}S\left( p\right) \right) $.\\

\end{center}

\begin{proposition}
\label{ti-iprop}\textit{Consider the game }$G\left( n,t,p,k\right) ,$ where $%
t_{i}=i,$ $p_{i}$ is decreasing in $i$ and $k=n=2m+1.$ Then

\begin{enumerate}
\item An optimal strategy for the searcher is to choose the set $\left\{
j,n-j\right\} $ with probability $1/\left( p_{j~}S\left( p\right) \right) $
for $j=m+1,\dots ,n$.

\item An optimal strategy for the hider is to choose location $j$ with
probability $\bar{h}_j$
for $j\geq m+1$ and not to choose locations $j\leq m$ at all.

\item The value of the game is $V=\frac{1}{S(p)}$.\newline
\end{enumerate}
\end{proposition}

\textbf{Proof.} Suppose the searcher adopts the strategy suggested above.
Any location $i$ that the hider chooses belongs to one of the sets of the
form $\left\{ j,n-j\right\} $ for $j=m+1,\dots ,n,$ where the set $\left\{
n,0\right\} $ denotes the set $\left\{ n\right\} .$ Since for $j\geq m+1$ we
have $j>n-j$ and the $p_{i}$ are decreasing, the hider is better off
choosing location $j.$ In this case he is found with probability $1/\left(
p_{j~}S\left( p\right) \right) $ and hence he is captured with probability
at least $p_{j}$ $\left( 1/\left( p_{j~}S\left( p\right) \right) \right)
=1/S\left( p\right) .$

Suppose the hider adopts the hiding distribution suggested above. Note that
no pure search strategy can inspect more than one of the locations $j\geq
m+1.$ Suppose that location $j$ is inspected. Then the probability that the
searcher finds and captures the hider is given by $\bar{h}_j
 p_{j}=1/\left( p_{j}S\left( p\right) \right) $ $p_{j}=1/S\left(
p\right) .$ It follows that $S\left( p\right) $ is the value of the game.\\

It is natural to also analyse if Proposition 4 still holds true for $k=n=$ even number. For the simplicity of our notation and better readability of Proposition 4, we decided to write this separate section for even number. In the case where $k=n=$ even, the solution is exactly the same as their odd counterpart. More specifically $k = n = 2m$ has the same value and optimal strategies as $k=n=2m+1$. However, it is important to note that in the even case, both the searcher's and hider's optimal strategy is unique. For instance, $k=n=4$ has the same value and optimal strategies as $k=n=5$. The same can be said for 6 and 7, 8 and 9, etc. 

\begin{corollary}
\label{uniqueness cor}Assuming the $p_{i}$ are strictly decreasing in $i,$
the hider strategy $\bar{h}$ given above is uniquely optimal, but the
searcher strategy is not.
\end{corollary}

\begin{proof}
Let $h^{\ast }\neq \bar{h}$ be a hiding distribution. We must have $h^{\ast
}_j +h^{\ast }_{n-j} >\bar{h}_j +\bar{h}_{n-j} =1/\left( p_{j}S\left( p\right) \right) $ for some $j\geq
m+1;$ otherwise the total probability given by $h^{\ast }$ would be less
than $1.$ Against such a distribution $h^{\ast },$ suppose that the searcher
inspects the two locations $j$ and $n-j.$ Then the probability that the
searcher wins is given by $p_{j}h^{\ast }_j +p_{n-j}h^{\ast
}_{n-j} \geq p_{j}\left( h^{\ast }_j +h^{\ast
}_{n-j} \right) $ because $p_{j}<p_{n-j}.$ But by our previous
estimate $h^{\ast }_j +h^{\ast }_{n-j} >1/\left(
p_{j}S\left( p\right) \right) $ this means the searcher wins with
probability at least $p_{j}\left( 1/\left( p_{j}S\left( p\right) \right)
\right) =1/S\left( p\right) $ and hence $h^{\ast }$ is not optimal.

Next, consider the searcher strategy which gives the same probability as
above for all the sets $\left\{ j,n-j\right\} $ for $j\geq m+2$ but gives
some of the probability assigned to $\left\{ m+1,m\right\} $ to the set $%
\left\{ m+1,m-1\right\} .$ Let's say the probability of $\left\{
m+1,m-1\right\} $ is a small positive number $\varepsilon .$ The total
probability of inspecting location $m+1$ (and all larger locations) has not
changed. The probability of inspecting location $m$ has gone down by $%
\varepsilon $. So the only way the new searcher strategy could fail to be
optimal is potentially when the hider chooses location $m$. In this case the
probability that the searcher wins is given by%
\begin{equation*}
\left( \left( 1/\left( p_{m+1~}S\left( p\right) \right) \right) -\varepsilon
\right) ~p_{m}.
\end{equation*}%
Comparing this to the value of the game, we consider the difference%
\begin{equation*}
\left( \left( 1/\left( p_{m+1~}S\left( p\right) \right) \right) -\varepsilon
\right) ~p_{m}-\frac{1}{S\left( p\right) }=\frac{p_{m}-p_{m-1}}{%
p_{m}~S\left( p\right) }-\varepsilon p_{m}.
\end{equation*}%
Since the first term on the right is positive because $p_{m}>p_{m-1},$ the
difference will be positive for sufficiently small positive $\varepsilon .$
\end{proof}

We will now consider an example to show how the solution changes as $k$ goes
up from the solved case of $k=n$. We conjecture that there exist a threshold with respect to $k$ in which above that threshold, the hider
's optimal strategy is to hide at location $n$. To determine that threshold we use the following idea.

\begin{proposition}
\label{max k prop}The game $G\left( n,p,t,k\right) $ has value $v=p_{n}$ if
and only if the value $v^{\prime }$ of the game $G\left( n-1,\left(
p_{1},\dots ,p_{n-1}\right) ,\left( 1,2,\dots ,n-1\right) ,k-n\right) $
(with the last location removed and resources reduced by $n$) is at least $%
p_{n}.$
\end{proposition}

\begin{proof}
Suppose $v=p_{n}.$ Every search set with positive probability must include
location $n,$ otherwise simply hiding there implies $v<p_{n}.$ So the
remaining part of every search set has $k^{\prime }=k-n.$ With this amount
of resources, the searcher must find the hider in the first $n$ locations
with probability at least $p_{n},$ which is what is stated in the
Proposition. Otherwise, the searcher will either have to not search location 
$n$ certainly (which gives $v<p_{n}$) or not search the remaining locations
with enough resources to ensure $v\geq p_{n}.$
\end{proof}

\subsection{An example with $k=10, n=5$.}
Consider the example where $p=\left( .5,.4,.3,.2,.1\right) $ with $%
k=10,n=5.$ Here $p_{n}=.1$ . The game with $p^{\prime }=\left(
.5,.4,.3,.2\right) $ and $k^{\prime }=k-n=5$ has value at least $.1$ because
of the equiprobable search strategy of $\left\{ 1,4\right\} $ and $\left\{
2,3\right\} .$ Here each location in the new game is inspected with the same
probability 1/2 and consequently the best the hider can do is to hide in the
best location 4, and then the searcher wins with probability $\left(
1/2\right) \left( .2\right) =.1.$ It follows from Proposition \ref{max k
prop} that the original game has the minimum possible value of $%
v=p_{n}=p_{5}=.1$ .

\subsection{Illustrative examples }

In this section we will use an example to further illustrate Proposition \ref{ti-iprop} and Corollary \ref{uniqueness cor}. We will also discuss the case when $k=n=$ even number.

First, we consider the game where $k=n=5,$ $t_{i}=i,$ and $p=\left(
.5,.4,.3,.2,.1\right) .$ The game matrix, excluding dominated search
strategies, is given by 
\begin{equation*}
\begin{tabular}{llllll}
A\TEXTsymbol{\backslash}location & 1 & 2 & 3 & 4 & 5 \\ \cline{2-6}
$\left\{ 5\right\} $ & \multicolumn{1}{|l}{0} & \multicolumn{1}{|l}{0} & 
\multicolumn{1}{|l}{0} & \multicolumn{1}{|l}{0} & \multicolumn{1}{|l|}{.1}
\\ \cline{2-6}
$\left\{ 1,4\right\} $ & \multicolumn{1}{|l}{.5} & \multicolumn{1}{|l}{0} & 
\multicolumn{1}{|l}{0} & \multicolumn{1}{|l}{.2} & \multicolumn{1}{|l|}{0}
\\ \cline{2-6}
$\left\{ 2,3\right\} $ & \multicolumn{1}{|l}{0} & \multicolumn{1}{|l}{.4} & 
\multicolumn{1}{|l}{.3} & \multicolumn{1}{|l}{0} & \multicolumn{1}{|l|}{0}
\\ \cline{2-6}
$\left\{ 1,3\right\} $ & \multicolumn{1}{|l}{.5} & \multicolumn{1}{|l}{0} & 
\multicolumn{1}{|l}{.3} & \multicolumn{1}{|l}{0} & \multicolumn{1}{|l|}{0}
\\ \cline{2-6}
$\left\{ 1,2\right\} $ & \multicolumn{1}{|l}{.5} & \multicolumn{1}{|l}{.4} & 
\multicolumn{1}{|l}{0} & \multicolumn{1}{|l}{0} & \multicolumn{1}{|l|}{0} \\ 
\cline{2-6}
\end{tabular}%
\end{equation*}%
The unique solution for the optimal hiding distribution is $\left(
0,0,2/11,3/11,6/11\right) $ and the value is $6/110=1/\left(
1/.3+1/.2+1/.1\right) \simeq .055$ . The optimal search strategy mentioned
in Proposition \ref{ti-iprop} is to play $\left\{ 5\right\} ,\left\{
1,4\right\} $ and $\left\{ 2,3\right\} $ with respective probabilities $%
6/11, $ $3/11$ and $2/11.$ Another strategy is to play $\left\{ 5\right\} $
and $\left\{ 1,4\right\} $ the same but to play $\left\{ 2,3\right\} $ and $%
\left\{ 1,3\right\} $ with probabilities $3/22$ and $1/22.$ It is of
interest to see how the solution of the game changes when $k$ increases from 
$k=n=5$ to higher values. We know that we need go no higher than $k=10$ from
Proposition \ref{max k prop} because in the game on locations $1$ to $4$
with $k^{\prime }=10-5=5,$ the searcher can inspect $\left\{ 4,1\right\} $
with probability $2/3$ and $\left\{ 3,2\right\} $ with probability $1/3$ to
ensure winning with probability at least $1/10=p_{5}.$

So we know the solution of the game for $k=5$ and $k\geq 10.$ The following
table gives the value of the game and the unique optimal hiding distribution
for these and intermediate values. (The optimal search strategies are varied
and we don't list them, though they are easily calculated.)%
\begin{gather*}
\begin{tabular}{lllllllll}
$k\backslash i$ &  & $1$ & $2$ & $3$ & $4$ & $5$ &  & Value \\ 
$5$ &  & 0 & 0 & 2/11 & 3/11 & 6/11 &  & $3/55~~~\simeq 0.0545$ \\ 
$6$ &  & 0 & 0 & 2/11 & 3/11 & 6/11 &  & $3/55~~~\simeq 0.0545$ \\ 
$6$ &  & 0 & 0 & 0 & 1/3 & 2/3 &  & $1/15~~~\simeq 0.06\,67$ \\ 
$8$ &  & 0 & 0 & 0 & 1/3 & 2/3 &  & $1/15~~~\simeq 0.06\,67$ \\ 
$9$ &  & 0 & 3/37 & 4/37 & 6/37 & 24/37 &  & $18/185\simeq 0.0943$ \\ 
$\geq 10$ &  & 0 & 0 & 0 & 0 & 1 &  & $1/10~~~=0.1$%
\end{tabular}
\\
\text{Table 1. Optimal hiding distribution and values, }k\geq 5.
\end{gather*}%
We know that the value must be nondecreasing in $k,$ but we see that it is
not strictly increasing. Roughly speaking (but not precisely), the hider
restricts towards fewer and better locations as $k$ increases, staying
always at the best location 5 for $k\geq 10.$ However there is the anomalous
distribution for $k=9$ which includes sometime hiding at location 2.

\subsection{Game with two types of locations}

In this section we analyse a more specific scenario where all available hiding locations are of two types. This model might be vaguely applied to military practices. Suppose a team of law enforcement is to capture a hiding fugitive in an apartment complex. Then all possible hiding locations can be reduced to a number of types, e.g. smaller rooms have similar shorter search times and higher capture probability than a parking lot. Here we solve the resulting search-pursuit game. 

Suppose there are two types of locations (hiding places). Type 1 takes time $%
t_{1}$ $=1$ (this is a normalization) to search, while type 2 takes time $%
t_{2}=\tau $ to search, with $\tau $ being an integer. Now let type 1
locations have capture probability $p$ while type 2 locations have capture
probability $q$. Moreover, suppose there are $a$ locations of type 1 and $b$
locations of type 2. The searcher has total search time $k.$ To simplify our
results we assume that $k$ is small enough such that $a\geq k$ (the searcher
can restrict all his searches to type 1) and $b\tau \geq k$ (he can also
restrict all his searches to type 2 locations).

Let $m=\lfloor k/\tau \rfloor $ be the maximum number of type 2 locations
that can be searched. The searcher's strategies are to search $j=0,1,\dots
,m $ type 2 locations (and hence $k-\tau j$ locations of type 1). Since all
locations of a given type are essentially the same, the decision for the
hider is simply the probability $y$ to hide at a randomly chosen location of
type 1 (and hence hide at a randomly chosen location of type 2 with
probability $1-y).$

Then the probability $P\left( j,y\right) $ that the searcher wins the game
is given by

\begin{align*}
& yp(\frac{k-\tau j}{a})-(1-y)q(\frac{j}{b}) \\
& =\frac{k}{a}py+\Big(\frac{q}{b}(1-y)-\frac{1}{a}py\tau \Big)j
\end{align*}

This will be independent of the searcher's strategy $j$ if

\begin{align*}
\frac{q}{b}(1-y)-\frac{1}{a}py\tau & =0,\text{or } \\
y& =\bar{y}\equiv \frac{aq}{aq+bp\tau }
\end{align*}

For $y = \bar{y}$, the capture probability is given by

\begin{equation*}
P(j,\bar{y})=\frac{pqk}{aq+bp\tau }
\end{equation*}

By playing $y = \bar{y}$, the hider ensures that the capture probability
(payoff) does not exceed $P(j,\bar{y})$.\newline

We now consider how to optimize the searcher's strategy.\textit{\ } Suppose
the searcher searches $j$ locations of type 2 with probability $x_{j}$, $%
j=0,1,\dots ,m$. If the hider is at a type 2 location then he is captured
with probability

\begin{align*}
\sum_{j=0}^{m}x_{j}\frac{qj}{b}& =\frac{q}{b}\sum_{j=0}^{m}jx_{j}=\frac{q}{b}%
\hat{\jmath},\text{ where} \\
\hat{\jmath}& =\sum_{j=0}^{m}jx_{j}
\end{align*}%
is the mean number of searches at type 2 locations. Similarly, if the hider
is at a type 1 location, the hider is captured with probability

\begin{align*}
\sum_{j=0}^{m}x_{j}\frac{p(k-\tau j)}{a}& =\frac{pk}{a}-\frac{p\tau }{a}%
\sum_{j=0}^{m}jx_{j} \\
& =\frac{pk}{a}-\frac{p\tau }{a}\hat{\jmath}
\end{align*}

It follows that the capture probability will be the same for hiding at
either location if we have

\begin{align*}
\frac{q}{b}\hat{\jmath}& =\frac{pk}{a}-\frac{p\tau }{a}\hat{\jmath},\text{
or,} \\
\hat{\jmath}& =\frac{pbk}{bp\tau +aq}.
\end{align*}
So for any probability distribution over the pure strategies $j\in $ $%
\{0,1,\dots ,m\}$ with mean $\hat{\jmath}$, the probability of capturing a
hider located either at a type 1 or a type 2 location is given by

\begin{equation*}
\frac{q}{b}\hat{\jmath}=\frac{pk}{a}-\frac{p\tau }{a}\hat{\jmath}=\frac{pqk}{%
aq+bp\tau }
\end{equation*}
To summarize, we have shown the following.

\begin{proposition}
\label{two types prop}\textit{Suppose all the hiding locations are of two
types: }$\mathit{a}$\textit{\ locations of type 1 with search time 1 and
capture probability $p$; }$\mathit{b}$\textit{\ locations of type 2 with
search time $\tau $ and capture probability $q$. Suppose }$\mathit{a}$%
\textit{\ and }$\mathit{b}$\textit{\ are large enough so the searcher can do
all his searching at a single location type, that is, $k\leq max(a,\tau b)$.
Then the unique optimal strategy for the hider is to hide in a random type
1 location with probability $\bar{y}=\frac{aq}{aq+bp\tau }$ and in a random
type 2 location with probability $1-\bar{y}$. Note that this is independent
of $k$. A strategy for the searcher which inspects $j$ locations of type 2
(and thus, $k-j\tau $ for type 1) with probability $x_{j}$ is optimal if and
only if the mean number }$\mathit{\hat{\jmath}=\sum_{j=0}^{m}jx_{j}}$\textit{%
, $m=\lfloor k/\tau \rfloor $ of type 2 locations inspected is given by $%
\hat{\jmath}=\frac{pbk}{bp\tau +aq}$. If this number is an integer, then the
searcher has an optimal pure strategy. The value of the game is given by $%
\frac{pqk}{aq+bp\tau }$}.
\end{proposition}

\section{Game Where Capture Probabilities are Unknown But Learned}

In this section we determine how the players can \textit{learn} the values
of the capture probabilities over time, starting with some \textit{a priori}
values and increasing these at locations from which there have been
successful escapes. This of course requires that the game is repeated. Here
we consider the simplest model, just two rounds. So after a successful
escape in the second round, we consider that the hider-prey has won the game
(Payoff $0$). More rounds of repeated play are considered in Gal,
Alpern,Casas (2015), but learning is not considered there.

We begin our analysis with two hiding locations, one of which may be
searched in each of the two rounds. If the hider is found at location $i$, 
he is captured with a probability $1-q_{i}$ (escapes with complementary
probability $q_{i}$). There are two rounds. If the hider is not found
(searcher looks in the wrong location) in either round, he wins and the
payoff is 0: If the hider is found and captured in either round, the
searcher wins and the payoff is 1: If the Hider is found but escapes in the
first round, the game is played one more time and both players remember
which location the hider escaped from. If the hider escapes in the second
(final) round, he wins and the payoff is 0.

The novel feature here is that the capture probabilities must be learned
over time. At each location, the capture probability is chosen by Nature
before the start of the game, independently with probability 1/2 of being $h$
(high) and probability 1/2 and being $l$ (the low probability), with $h>l$.
In the biological scenario, this may be the general distribution of
locations in a larger region in which it is easy or hard to escape from. A
more general distribution is possible within our model, but this two point
distribution is very easy to understand. If there is escape from location $i$
in the first round, then in the second round the probability that the
capture probability at $i$ is $h$ goes down (to some value less than $1/2$).
This is a type of Bayesian learning, which only takes place after an escape,
and only at the location of the escape.

Our model contributes to the realistic interaction between searching-predator and hiding-prey acting in a possibly changing environment. Most often in nature, the searcher has no or incomplete information during the search and pursuit interaction. particularly in Mech, Smith, and MacNulty (2015), a pack of wolves has to learn over time the difficulty of pursuing their prey in specific terrain. Moreover, hiding-prey such as elk seems to prefer areas with lots toppled dead trees, creating an entanglement of logs difficult to travel through. We focus here on asking questions if learning the capture probabilities will affect the searching and hiding behaviour. More specifically, suppose an elk manages to escape through the deep forest, should it stay there where he believes the capture probability is low enough, or hide at a different location?

\subsection{Normal form of the two-period learning game}

We use the \textit{normal form} approach, rather than a repeated game
approach. A strategy for either player says where he will search/hide in the
two periods (assuming the game goes to the second period). Due to the
symmetry of the two locations, both players cannot but choose their first
period search or hide locations randomly. Thus the players have two
strategies: $rs$ (random,same) and $rd$ (random,different). If there is a
successful escape from that location, they can either locate in the same
location (strategy $rs$) or the other location (strategy $rd)$. This gives a
simple two by two matrix game. In this subsection we calculate its normal
form; in the next subsection we present the game solution. \newline

First we compute the payoff for the strategy pair $(rs,rs)$: Half the time
both players (searcher and hider) go to different locations in first period, in which case the hider wins
and the payoff is 0. So we ignore this, put in a factor of (1/2), and assume
they go to the same location in the first period. There is only one location
to consider, suppose it has escape probability $x$. Then, as they both go
back to this location in the second period if the hider escapes in the first
period, the expected payoff is given by

\begin{equation}
P_{x}(rs,rs)=(1/2)~~((1-x)1+x(1-x))\newline
\label{Px}.
\end{equation}%
Since $x$ takes values $l$ and $h$ equiprobably we have

\begin{eqnarray}
P(rs,rs) &=&\frac{P_{h}(rs,rs)+P_{l}(rs,rs)}{2}  \notag \\
&=&\frac{2-h^{2}-l^{2}}{4}  \label{Prsrs}.
\end{eqnarray}%
It is worth noting two special cases: If both escape probabilities are 1
(escape is certain), then the hider always wins and the payoff is $0.$ If
both escape probabilities are $0$ then the searcher wins if and only if they
both choose the same location, which has probability $1/2.$

Next we consider the strategy pair $(rd,rd)$. Here we can assume they both
go to location 1 in the first period (hence we add the factor of 1/2) and
location 2 in the second period. The escape probabilities at these ordered
locations 1 and 2 can be any of the following: $hh,ll,hl,lh$. The first two
are straightforward as it is the same as going to the same location twice
(already calculated in (\ref{Prsrs})). We list the calculation of the four
ordered hiding locations below, where $P_{x}$ is given in (\ref{Px}).

\begin{align*}
P_{hh}(rd,rd)& =P_{h}(rs,rs) \\
P_{ll}(rd,rd)& =P_{l}(rs,rs) \\
P_{lh}(rd,rd)& =(1/2)((1-l)1+l(1-h)) \\
P_{hl}(rd,rd)& =(1/2)((1-h)1+h(1-l))
\end{align*}%
Taking the average of these four values gives,

\begin{equation}
P(rd,rd)=\frac{4-h^{2}-l^{2}-2hl}{8}=\frac{4-(h+l)^{2}}{8}.  \label{rd,rd}
\end{equation}%
\newline
Now consider the strategy pair $(rs,rd)$. If they go to different locations
in the first period, the game ends with payoff 0. So again, we put in factor
of 1/2 and assume they go to same location in first period. This means that
if an escape happens in the first period, the hider wins (payoff 0) in the
second period. So the probability the searcher wins is

\begin{align}
P(rs,rd)& =P(rd,rs)=(1/2)\Big(1/2((1-h)+(1-l))\Big)  \notag \\
& =\frac{2-(h+l)}{4}  \label{rs.rd}.
\end{align}%
\newline
Thus, we have completed the necessary calculations and the game matrix for
for the strategy pairs $rs$ and $rd$, with searcher as the maximizer.

To solve this game, we begin with the game matrix as follows;

\begin{align*}
A& =A(l,h)=\left[ 
\begin{array}{cc}
P\left( rs,rs\right) & P\left( rs,rd\right) \\ 
P\left( rd,rs\right) & P\left( rd,rd\right)%
\end{array}%
\right] \\
& =%
\begin{bmatrix}
\frac{2-(h^{2}+l^{2})}{4} & \frac{2-(h+l)}{4} \\ 
\frac{2-(h+l)}{4} & \frac{4-(h+l)^{2}}{8}%
\end{bmatrix}%
\end{align*}

Then we take out the fraction 1/8 to the left hand side of the equation, and we have\\

\begin{equation*}
8A =%
\begin{bmatrix}
-2h^{2}-2l^{2}+4 & 4-2h-2l \\ 
4-2h-2l & 4-(h+l)^{2}%
\end{bmatrix}
\end{equation*}

At this point we try to make the right-hand side of the equation to be a diagonal matrix so we can easily compute it. Therefore we can write the equation as follow

\begin{equation*}
8A-(4-2h-2l)%
\begin{bmatrix}
1 & 1 \\ 
1 & 1%
\end{bmatrix}%
 =Y=%
\begin{bmatrix}
-2h^{2}+2h-2l^{2}+2l & 0 \\ 
0 & 2h+2l-(h+l)^{2}%
\end{bmatrix}%
. \\
\end{equation*}

Note that $V(A)$ is the value of the matrix A. From the equation above, it shows that the right-hand side of the equation is a diagonal matrix, and a simple formula for the value of diagonal matrix games is as follow

\begin{equation*}
V\left( 
\begin{bmatrix}
a & 0 \\ 
0 & b%
\end{bmatrix}%
\right) =1/\left( 1/a+1/b\right) .
\end{equation*}

Using the above formula, we have 

\begin{equation*}
V\left( 8A-(4-2h-2l)%
\begin{bmatrix}
1 & 1 \\ 
1 & 1%
\end{bmatrix}%
\right)  =V\left( Y\right) =\frac{1}{\frac{1}{-2h^{2}+2h-2l^{2}+2l}+\frac{1%
	}{2h+2l-(h+l)^{2}}}.
\end{equation*}

Computing this for the value of game matrix A, we have the following equation for $V(A)$,

\begin{equation}
V\left( A\right) =\frac{1}{2}-\frac{1}{4}l-\allowbreak \frac{1}{4}h-\frac{1}{%
	8\left( \frac{1}{2h^{2}-2h+2l^{2}-2l}-\frac{1}{2h+2l-\left( h+l\right) ^{2}}%
	\right) }.
\end{equation}

It is also important to note that in a diagonal game, players adopt each strategy with a probability inversely proportional to its diagonal element. To obtain this we first calculate the value of $V(Y)$ given above. Then, both the searcher and hider should choose $rs$ and $rd$ with probabilities  $V(Y)/a$ and  $V(Y)/b$ respectively.\\

We can now see that, as expected, a successful escape from a location makes
that location more attractive to the hider as a future hiding place. This is
confirmed in the following.

\begin{proposition}
In the learning game when $l<h,$ after a successful escape both players
should go back to the same location with probability greater than 1/2.
\end{proposition}

\begin{proof}
Let $a$ and $b$ denote, as above, the diagonal elements of $Y.$ We have%
\begin{eqnarray*}
a-b &=&\left( -2h^{2}+2h-2l^{2}+2l\right) -\left( 2h+2l-(h+l)^{2}\right) \\
&=&-\left( h-l\right) ^{2}<0.
\end{eqnarray*}%
This means that $b>a$ and $V/a>V/b.$ Hence by the observation (5)
the strategy $rs$ should be played with a higher probability $\left(
V/a\right) $ than $rd$ (probability $V/b),$ in particular with probability
more than $1/2.$
\end{proof}

\subsection{An example with $l = 1/3$ and $h= 2/3$}

A simple example is when the low escape probability is $l=1/3$ and the high
escape probability is $h=2/3$: This give the matrix $A$ as\newline

\begin{equation*}
\centering A(l,h)=%
\begin{bmatrix}
\frac{13}{36} & \frac{1}{4} \\ 
\frac{1}{4} & \frac{3}{8}%
\end{bmatrix}%
\end{equation*}%
\newline
with value $V=V(1/3,2/3)=21/68$, and where each of the player optimally
plays $rs$ with probability $9/17$ and $rd$ with probability $8/17$. \newline

Suppose there is an escape in the first period at say location 1. Then in
the second period the hider goes to location 1 with probability 9/17. Since
the subjective probability of capture at location 2, from the point of view
of either player, remains unchanged at (1/3 + 2/3) /2 = 1/2; this
corresponds to a certain probability $x$ at location 1, that is, a matrix

\begin{equation*}
\centering%
\begin{bmatrix}
x & 0 \\ 
0 & 1/2%
\end{bmatrix}%
\end{equation*}%
\newline
We then have that \newline

\begin{align*}
(9/17)x& =(8/17)(1/2)\text{ or, } \\
x& =4/9.
\end{align*}%
This corresponds to the probability of escape probability $l=1/3$ of $q$,
where

\begin{align*}
\centering q1/3+(1-q)2/3& =4/9\text{ or, } \\
q& =2/3.
\end{align*}

Thus, based on the escape at location 1 in the first period, the
probability that the escape probability there is 1/3 has gone up from the
initial value of 1/2 to the higher value of 2/3.

\section{ Summary}

The breakthrough paper of Gal and Casas (2014) gave us a model in which both
the search and pursuit elements of predator-prey interactions could be
modeled together in a single game. In that paper the capture probabilities
depended on the hiding location but the time required to search a location
was assumed to be constant. In the first part of this paper, we drop that
simplifying assumption. We first consider a particular scenario where we order the locations such that the search times
increase while the capture probabilities decrease. We solve
this game for the case of a particular total search time of the searcher. We
then consider a scenario where there are many hiding locations but they come
in only two types. Locations of each type are identical in that they have
the same search times and the same capture probabilities. We solve the
resulting search-pursuit game.

In the second part of the paper we deal with the question of how the players
(searcher-predator and hider-prey) learn the capture probabilities of the
different locations over time. We adopt a simple Bayesian approach. After a
successful escape from a given location, both players update their
subjective probabilities that it is a location with low or high capture
probability; the probability that it is low obviously increases. In the game
formulation, the players incorporate into their plan the knowledge that if
there is an escape, then that location becomes more favorable to the hider
in the next period.\\

The search-hide and pursuit-evasion game is quite difficult and finding a solution for the most general case is quite challenging. Most probably, it is a good idea for the next step to solve for a more specific question in the problem. 

We consider a possible extension to Proposition 7 by analysing larger $k$ . Consider the example $a = b = 1;t_1 = 1;t_2 = 3;k = 4$; and say $p < q $ ($p_1 < p_2$ as in Gal- Casas(2014)). The Searcher inspects both cells (one of each type), so he certainly finds the Hider. He captures him with probability $p$ if the hider is at location 1, and $q$ if at location type 2. So the Hider should hide at location of type 2 as it has lower capture probability. The main question will be: How big does $k$ have to be for this to occur? And are there only two solution types as in Gal-Casas (2014)? We conjecture that, as in Gal-Casas, there is a critical value of $k = \hat{k}$ such that for $k < \hat{k}$, Proposition 7 applies, and for $k \geq \hat{k}$ the Hider locates in a cell of the type with the lower capture probability.

The game with learning model has also been analysed using dynamic form (Alpern, Gal, Lee, and Casas (2019)). This allows more effective analysis for more than two locations and two rounds. Moreover, we believe the next avenue of research is to consider the non-zero-sum game. Indeed, one may argue that a game between a predator and a prey may not necessarily be a zero-sum game, as the predator is hunting it’s dinner while the prey is running for survival. This is an important if challenging aspect to deal with for future studies.
\\

\bigskip Authors' Address: Warwick Business School, University of Warwick,
Coventry CV4 7AL, United Kingdom\newline
\qquad steve.alpern@wbs.ac.uk (+44) 7971 854148; phd16vl@mail.wbs.ac.uk

\end{document}